\documentclass[fleqn,twoside]{article}
\usepackage{espcrc1}

\newcommand{\AmS}{{\protect\the\textfont2
  A\kern-.1667em\lower.5ex\hbox{M}\kern-.125emS}}
\title{Towards Decision Support Technology Platform  for Modular Systems}

\author{Mark Sh. Levin
\thanks{
 Mark Sh. Levin:~
 Inst. for Inform. Transmission Problems, Russian Academy of
 Sciences;
 http://www.mslevin.iitp.ru;
 email: mslevin@acm.org
  }
  }

\begin{document}

\maketitle

\begin{abstract}
 The survey methodological paper addresses
 a glance to a general decision support platform technology
 for modular systems
 (modular/composite alterantives/solutions)
  in various applied domains.
 The decision support platform consists of seven basic
 combinatorial engineering frameworks
 (system synthesis, system modeling, evaluation,
 detection of bottleneck, improvement/extension,
 multistage design, combinatorial evolution and forecasting).
 The decision support platform is based on decision support
 procedures
 (e.g., multicriteria selection/sorting, clustering),
  combinatorial optimization problems
  (e.g., knapsack, multiple choice problem, clique,
  assignment/allocation, covering, spanning trees),
 and their combinations.
 The following is described:
 (1) general scheme of the decision support platform technology;
 (2) brief descriptions of modular (composite) systems (or composite
 alternatives);
 (3) trends in moving from chocie/selection of alternatives to
 processing of composite alternatives which correspond to
 hierarchical modular products/systems;
 (4) scheme of resource requirements
 (i.e., human, information-computer);
 and
 (5) basic combinatorial engineering frameworks and their applications
 in various domains.

~~~~~~~~~~~


\vspace{1pc}
\end{abstract}


\newcounter{cms}
\setlength{\unitlength}{1mm}

\section{Introduction}

 In recent years the significance of
 modular products/systems and corresponding product families
 (or product lines)
 has been increased
 (e.g., \cite{dahmus01,du01,du02,jiao07,simp01,simp06,simp14}).
 Some basic research directions in the fields of modularity and
 modular systems are briefly pointed out in Table 1
 (e.g.,
 mechanical systems,
 manufacturing systems, robots, software systems, computing systems, electronic systems, Web-based systems,
 communication protocols, control systems).

\begin{center}
{\bf Table 1.} Basic research directions in modularity/modular systems \\
\begin{tabular}{| l  l  | l |}
\hline

  &Research direction & Some sources \\
\hline

 1.&Modularity               &\cite{agerfalk07,ali02,bald00,cai05,ethiraj04,garud09,huang98,martin02} \\
 2.&Modular products/systems &\cite{ber93,dahmus01,ger99,huang98,jiao07,kamrani00,lev98,lev06,lev15,navabi07}\\
 3.&Modularity and commonality research &\cite{cai12,fel06,fix07,fujita02,fujita13,ger88,lev06}\\
 4.&Products/systems configuration
 &\cite{conradi98,corbett04,fru04,helo10,lev98,lev06,lev09,lev12morph,lev12a,lev15,mcd82,mck04,sabin98,soi98,song11,spicer02,tam02,ta11,wielinga97,yang08}\\
 5.&Reconfiguration, reconfigurable systems&\cite{bi08,bobda07,bon02,card11,comp02,elma09,fer08,koren99,lev98,lev06,lev09,lev12morph,lev13imp,lev15}\\
 6.&Adaptable design of products/systems &\cite{elma09,flet07,gu04,has05,kas07,li08,xue12}\\
 7.&Design of products/systems for variety&\cite{erens96,fujita02,fujita04,martin99,martin02,prasad98}\\
 8.&Product families         &\cite{corbett04,du01,fel06,fer08,fujita13,gon02,jiao07,lev06,simp06,simp14}\\
 9.&Product platforms        &\cite{corbett04,du02,gon00,gon02,jiao07,keut00,kim96,martin02,meyer97,rob98,simp01,simp06,simp14}\\
 10.&Approaches to general decision support &\cite{corbett04,fru04,lev13intro,lev15,sabin98,wielinga97,yang08}\\
  &platform &\\


\hline
\end{tabular}
\end{center}

 Fig. 1 depicts a traditional scheme of product platform efforts
  for a certain product domain
  (e.g., buildings, software,
  manufacturing systems,
  aerospace systems, ships,
  mechatronic  systems, computing systems, etc.)
 \cite{rob98,simp01,simp06,simp14}.


\begin{center}
\begin{picture}(94,36)

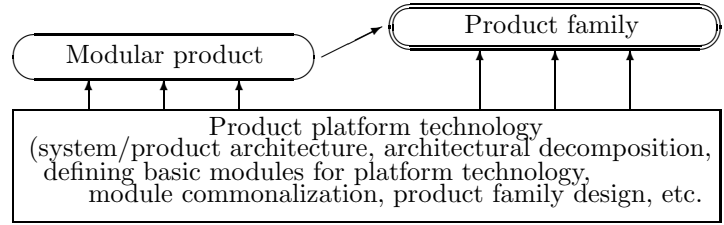
\put(02,00){\makebox(0,0)[bl]{Fig. 1. Traditional scheme of
 product platform technology}}


\put(20,28){\oval(40,06)}

\put(07,26.5){\makebox(0,0)[bl]{Modular product}}

\put(41,28){\vector(2,1){08}}

\put(72,32){\oval(44,06)} \put(72,32){\oval(43,05)}

\put(60,30.5){\makebox(0,0)[bl]{Product family}}


\put(62,21){\vector(0,1){8}}
\put(72,21){\vector(0,1){8}}\put(82,21){\vector(0,1){8}}


\put(10,21){\vector(0,1){4}}
\put(20,21){\vector(0,1){4}}\put(30,21){\vector(0,1){4}}


\put(00,06){\line(1,0){94}}\put(00,21){\line(1,0){94}}
\put(00,06){\line(0,1){15}} \put(94,06){\line(0,1){15}}

\put(26,17){\makebox(0,0)[bl]{Product platform technology}}
\put(02,14){\makebox(0,0)[bl]{(system/product architecture,
 architectural decomposition,}}

\put(04,11){\makebox(0,0)[bl]{defining basic modules for
 platform technology, }}

\put(10,08){\makebox(0,0)[bl]{module commonalization,
 product family design, etc. }}

\end{picture}
\end{center}

 Here,
 a general decision support platform technology is briefly described  that
 can be used for many engineering/management domains (Fig. 2)
 \cite{lev06,lev13intro,lev15}.


\begin{center}
\begin{picture}(102,59)

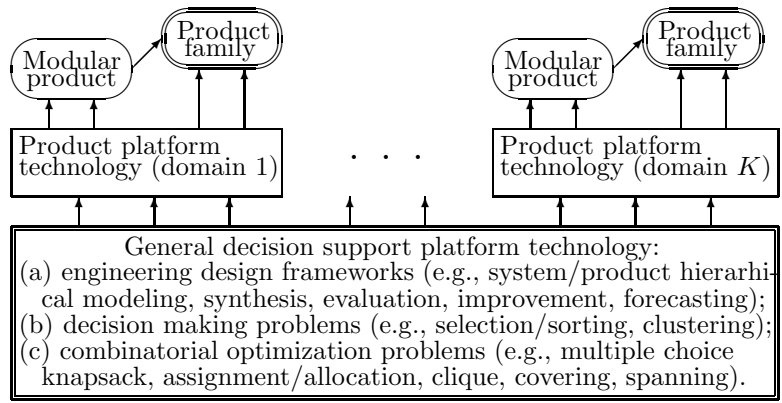
\put(02.6,00){\makebox(0,0)[bl]{Fig. 2. General decision
 support platform  for modular systems}}


\put(08,50){\oval(16,08)}

\put(01.5,50){\makebox(0,0)[bl]{Modular}}
\put(02,47){\makebox(0,0)[bl]{product}}

\put(16,50){\vector(1,1){04}}

\put(05,42){\vector(0,1){4}} \put(11,42){\vector(0,1){4}}



\put(28,54){\oval(16,08)} \put(28,54){\oval(15,07)}

\put(22,54){\makebox(0,0)[bl]{Product}}
\put(23,51){\makebox(0,0)[bl]{family}}

\put(25,42){\vector(0,1){8}} \put(31,42){\vector(0,1){8}}


\put(00,33){\line(1,0){36}}\put(00,42){\line(1,0){36}}
\put(00,33){\line(0,1){09}} \put(36,33){\line(0,1){09}}

\put(01,38){\makebox(0,0)[bl]{Product platform }}
\put(01,35){\makebox(0,0)[bl]{technology (domain \(1\))}}


\put(45,38){\makebox(0,0)[bl]{{\bf .~ .~ .}}}


\put(72,50){\oval(16,08)}

\put(65.5,50){\makebox(0,0)[bl]{Modular}}
\put(66,47){\makebox(0,0)[bl]{product}}

\put(80,50){\vector(1,1){04}}

\put(69,42){\vector(0,1){4}} \put(75,42){\vector(0,1){4}}



\put(92,54){\oval(16,08)} \put(92,54){\oval(15,07)}

\put(86,54){\makebox(0,0)[bl]{Product}}
\put(87,51){\makebox(0,0)[bl]{family}}

\put(89,42){\vector(0,1){8}} \put(95,42){\vector(0,1){8}}


\put(64,33){\line(1,0){38}}\put(64,42){\line(1,0){38}}
\put(64,33){\line(0,1){09}} \put(102,33){\line(0,1){09}}

\put(65,38){\makebox(0,0)[bl]{Product platform }}
\put(65,35){\makebox(0,0)[bl]{technology (domain \(K\))}}


\put(09,29){\vector(0,1){4}} \put(19,29){\vector(0,1){4}}
\put(29,29){\vector(0,1){4}}

\put(45,29){\vector(0,1){4}} \put(55,29){\vector(0,1){4}}

\put(73,29){\vector(0,1){4}} \put(83,29){\vector(0,1){4}}
\put(93,29){\vector(0,1){4}}

\put(00,06){\line(1,0){102}}\put(00,29){\line(1,0){102}}
\put(00,06){\line(0,1){023}} \put(102,06){\line(0,1){023}}

\put(0.5,06.5){\line(1,0){101}}\put(00.5,28.5){\line(1,0){101}}
\put(0.5,06.5){\line(0,1){022}} \put(101.5,06.5){\line(0,1){022}}

\put(15,24.5){\makebox(0,0)[bl]{General decision support platform
 technology:}}

 \put(01,21){\makebox(0,0)[bl]{(a) engineering design frameworks
 (e.g., system/product hierarhi- }}

 \put(04,17.5){\makebox(0,0)[bl]{cal modeling,  synthesis, evaluation, improvement, forecasting);}}

 \put(01,14){\makebox(0,0)[bl]{(b) decision making problems
 (e.g., selection/sorting, clustering);}}

 \put(01,10.5){\makebox(0,0)[bl]{(c) combinatorial optimization problems
  (e.g., multiple choice}}

 \put(04,07){\makebox(0,0)[bl]{knapsack, assignment/allocation, clique, covering,
  spanning).}}

\end{picture}
\end{center}


\begin{center}
\begin{picture}(114,99)

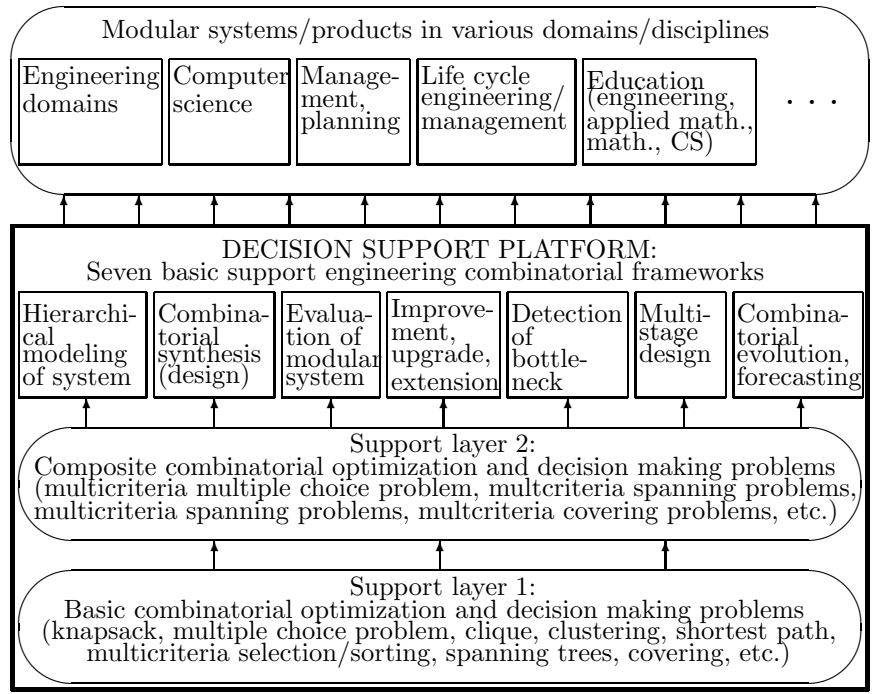
\put(08,00){\makebox(0,0)[bl]{Fig. 3. Scheme of general
 decision support platform technology}}


\put(01,76){\line(1,0){19}}\put(01,90){\line(1,0){19}}
\put(01,76){\line(0,1){14}} \put(20,76){\line(0,1){14}}

\put(1.5,86){\makebox(0,0)[bl]{Engineering}}
\put(1.5,83){\makebox(0,0)[bl]{domains}}
\put(1.5,80){\makebox(0,0)[bl]{}}
\put(1.5,77){\makebox(0,0)[bl]{}}


\put(21,76){\line(1,0){16}}\put(21,90){\line(1,0){16}}
\put(21,76){\line(0,1){14}} \put(37,76){\line(0,1){14}}

\put(21.5,86){\makebox(0,0)[bl]{Computer}}
\put(21.5,83){\makebox(0,0)[bl]{science}}
\put(21.5,80){\makebox(0,0)[bl]{}}
\put(21.5,77){\makebox(0,0)[bl]{}}


\put(38,76){\line(1,0){15}}\put(38,90){\line(1,0){15}}
\put(38,76){\line(0,1){14}} \put(53,76){\line(0,1){14}}

\put(38.5,86){\makebox(0,0)[bl]{Manage-}}
\put(38.5,83){\makebox(0,0)[bl]{ment,}}
\put(38.5,80){\makebox(0,0)[bl]{planning}}
\put(38.5,77){\makebox(0,0)[bl]{}}


\put(54,76){\line(1,0){21}}\put(54,90){\line(1,0){21}}
\put(54,76){\line(0,1){14}} \put(75,76){\line(0,1){14}}

\put(54.5,86){\makebox(0,0)[bl]{Life cycle}}
\put(54.5,83){\makebox(0,0)[bl]{engineering/}}
\put(54.5,80){\makebox(0,0)[bl]{management}}
\put(54.5,77){\makebox(0,0)[bl]{}}


\put(76,76){\line(1,0){23}}\put(76,90){\line(1,0){23}}
\put(76,76){\line(0,1){14}} \put(99,76){\line(0,1){14}}

\put(76.5,86){\makebox(0,0)[bl]{Education}}
\put(76.5,83){\makebox(0,0)[bl]{(engineering,}}
\put(76.5,80){\makebox(0,0)[bl]{applied math.,}}
\put(76.5,77){\makebox(0,0)[bl]{math., CS)}}

\put(103,84){\makebox(0,0)[bl]{{\bf . . .}}}


\put(12,92){\makebox(0,0)[bl]{Modular systems/products
 in various domains/disciplines}}

\put(57,84.5){\oval(114,25)}


\put(01,45){\line(1,0){17}}\put(01,59){\line(1,0){17}}
\put(01,45){\line(0,1){14}} \put(18,45){\line(0,1){14}}

\put(1.5,55){\makebox(0,0)[bl]{Hierarchi-}}
\put(1.5,52){\makebox(0,0)[bl]{cal }}
\put(1.5,49){\makebox(0,0)[bl]{modeling}}
\put(1.5,46){\makebox(0,0)[bl]{of system}}


\put(19,45){\line(1,0){16}}\put(19,59){\line(1,0){16}}
\put(19,45){\line(0,1){14}} \put(35,45){\line(0,1){14}}

\put(19.5,55){\makebox(0,0)[bl]{Combina-}}
\put(19.5,52){\makebox(0,0)[bl]{torial}}
\put(19.5,49){\makebox(0,0)[bl]{synthesis}}
\put(19.5,46){\makebox(0,0)[bl]{(design)}}


\put(36,45){\line(1,0){13}}\put(36,59){\line(1,0){13}}
\put(36,45){\line(0,1){14}} \put(49,45){\line(0,1){14}}

\put(36.5,55){\makebox(0,0)[bl]{Evalua-}}
\put(36.5,52){\makebox(0,0)[bl]{tion of}}
\put(36.5,49){\makebox(0,0)[bl]{modular}}
\put(36.5,46){\makebox(0,0)[bl]{system}}


\put(50,45){\line(1,0){15}}\put(50,59){\line(1,0){15}}
\put(50,45){\line(0,1){14}} \put(65,45){\line(0,1){14}}

\put(50.5,55){\makebox(0,0)[bl]{Improve-}}
\put(50.5,52){\makebox(0,0)[bl]{ment, }}
\put(50.5,49){\makebox(0,0)[bl]{upgrade, }}
\put(50.5,46){\makebox(0,0)[bl]{extension}}


\put(66,45){\line(1,0){16}}\put(66,59){\line(1,0){16}}
\put(66,45){\line(0,1){14}} \put(82,45){\line(0,1){14}}

\put(66.5,55){\makebox(0,0)[bl]{Detection}}
\put(66.5,52){\makebox(0,0)[bl]{of}}
\put(66.5,49){\makebox(0,0)[bl]{bottle-}}
\put(66.5,46){\makebox(0,0)[bl]{neck}}


\put(83,45){\line(1,0){12}}\put(83,59){\line(1,0){12}}
\put(83,45){\line(0,1){14}} \put(95,45){\line(0,1){14}}

\put(83.5,55){\makebox(0,0)[bl]{Multi-}}
\put(83.5,52){\makebox(0,0)[bl]{stage}}
\put(83.5,49){\makebox(0,0)[bl]{design}}
\put(83.5,46){\makebox(0,0)[bl]{}}


\put(96,45){\line(1,0){17.2}}\put(96,59){\line(1,0){17.2}}
\put(96,45){\line(0,1){14}} \put(113.2,45){\line(0,1){14}}

\put(96.5,55){\makebox(0,0)[bl]{Combina-}}
\put(96.5,52){\makebox(0,0)[bl]{torial}}
\put(96.5,49){\makebox(0,0)[bl]{evolution,}}
\put(96.5,46){\makebox(0,0)[bl]{forecasting}}


\put(07,68){\vector(0,1){4}} \put(17,68){\vector(0,1){4}}
\put(27,68){\vector(0,1){4}} \put(37,68){\vector(0,1){4}}
\put(47,68){\vector(0,1){4}}

\put(57,68){\vector(0,1){4}}

\put(67,68){\vector(0,1){4}} \put(77,68){\vector(0,1){4}}
\put(87,68){\vector(0,1){4}} \put(97,68){\vector(0,1){4}}
\put(107,68){\vector(0,1){4}}

\put(28,63.6){\makebox(0,0)[bl]{DECISION SUPPORT PLATFORM:
 }}

 \put(10,60){\makebox(0,0)[bl]{Seven basic support
 engineering combinatorial frameworks}}


\put(00,06){\line(1,0){114}}\put(00,68){\line(1,0){114}}
\put(00,06){\line(0,1){62}} \put(114,06){\line(0,1){62}}

\put(0.3,06.3){\line(1,0){113.4}}\put(0.3,67.7){\line(1,0){113.4}}
\put(0.3,06.3){\line(0,1){61.4}}
\put(113.7,06.3){\line(0,1){61.4}}


\put(10,41){\vector(0,1){4}}

\put(27,41){\vector(0,1){4}}

\put(42.5,41){\vector(0,1){4}}

\put(57.5,41){\vector(0,1){4}}

\put(74,41){\vector(0,1){4}}

\put(89.5,41){\vector(0,1){4}}

\put(105,41){\vector(0,1){4}}


\put(57,33.5){\oval(112,15)}

\put(45,37){\makebox(0,0)[bl]{Support layer 2: }}
\put(03,34){\makebox(0,0)[bl]{Composite combinatorial optimization
and decision making problems}}

\put(03,31){\makebox(0,0)[bl]{(multicriteria multiple choice
problem, multcriteria spanning problems,}}

\put(03,28){\makebox(0,0)[bl]{multicriteria spanning problems,
multcriteria covering problems, etc.)}}


\put(27,22){\vector(0,1){4}} \put(57,22){\vector(0,1){4}}
\put(87,22){\vector(0,1){4}}

\put(57,14.5){\oval(112,15)}

\put(45,18){\makebox(0,0)[bl]{Support layer 1: }}
\put(07,15){\makebox(0,0)[bl]{Basic combinatorial optimization and
decision making problems}}

\put(04,12){\makebox(0,0)[bl]{(knapsack, multiple choice problem,
 clique, clustering, shortest path,}}

\put(10,09){\makebox(0,0)[bl]{multicriteria selection/sorting,
 spanning trees, covering, etc.)}}

\end{picture}
\end{center}

\section{Scheme of General Decision Support Platform}

 A scheme of the proposed general decision support platform
 technology
 is shown in Fig. 3.
 Here, two support layers of decision making and
 combinatorial optimization problems/models are used:
 (i) basic problems,
 (ii) composite problems.

\section{Towards Hierarchical Modeling of Modular Systems}

 In general,
 knowledge representation in product design systems
 is systematically studied
 in \cite{chan13,hansen01}.
 Here, modular systems
 (or corresponding
 (modular/composite alterantives/solutions)
  are examined as the following
 (i.e., system configuration)
 (e.g., \cite{lev06,lev09,lev12hier,lev13intro,lev15}:

 (a) a set of system elements (components, modules),

 (b) a set of system elements and their interconnections
 (i.e., a special structure over the system elements, e.g.,
 hierarchy, tree-like structure).

 Fig. 4 depicts a composite (modular) system, consisting
 of \(n\) components/modules
 (and corresponding three design alternatives DAs for each
 component/module).

\begin{center}
\begin{picture}(100,45)
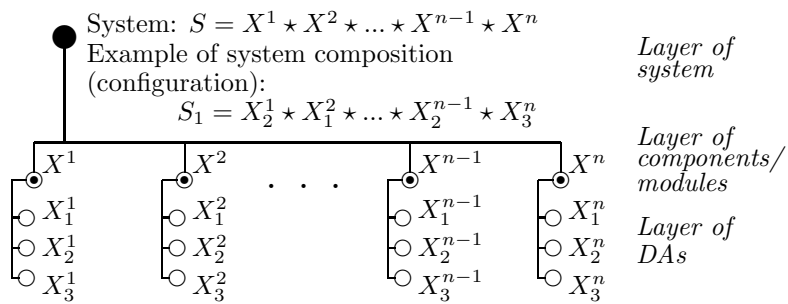
\put(08,00){\makebox(0,0)[bl] {Fig. 4. Illustration for composite
 (modular) system}}

\put(4,5){\makebox(0,8)[bl]{\(X^{1}_{3}\)}}
\put(4,10){\makebox(0,8)[bl]{\(X^{1}_{2}\)}}
\put(4,15){\makebox(0,8)[bl]{\(X^{1}_{1}\)}}

\put(24,5){\makebox(0,8)[bl]{\(X^{2}_{3}\)}}
\put(24,10){\makebox(0,8)[bl]{\(X^{2}_{2}\)}}
\put(24,15){\makebox(0,8)[bl]{\(X^{2}_{1}\)}}

\put(54,5){\makebox(0,8)[bl]{\(X^{n-1}_{3}\)}}
\put(54,10){\makebox(0,8)[bl]{\(X^{n-1}_{2}\)}}
\put(54,15){\makebox(0,8)[bl]{\(X^{n-1}_{1}\)}}

\put(74,5){\makebox(0,8)[bl]{\(X^{n}_{3}\)}}
\put(74,10){\makebox(0,8)[bl]{\(X^{n}_{2}\)}}
\put(74,15){\makebox(0,8)[bl]{\(X^{n}_{1}\)}}

\put(3,21){\circle*{1}} \put(23,21){\circle*{1}}
\put(53,21){\circle*{1}} \put(73,21){\circle*{1}}

\put(3,21){\circle{2}} \put(23,21){\circle{2}}
\put(53,21){\circle{2}} \put(73,21){\circle{2}}

\put(0,21){\line(1,0){02}} \put(20,21){\line(1,0){02}}
\put(50,21){\line(1,0){02}} \put(70,21){\line(1,0){02}}

\put(0,21){\line(0,-1){13}} \put(20,21){\line(0,-1){13}}
\put(50,21){\line(0,-1){13}} \put(70,21){\line(0,-1){13}}

\put(70,16){\line(1,0){01}} \put(70,12){\line(1,0){01}}
\put(70,8){\line(1,0){01}}

\put(72,16){\circle{2}}  \put(72,12){\circle{2}}
\put(72,8){\circle{2}}


\put(50,16){\line(1,0){01}} \put(50,12){\line(1,0){01}}
\put(50,8){\line(1,0){01}}

\put(52,16){\circle{2}} \put(52,12){\circle{2}}
\put(52,8){\circle{2}}


\put(20,8){\line(1,0){01}} \put(20,12){\line(1,0){01}}
\put(20,16){\line(1,0){01}}

\put(22,12){\circle{2}} \put(22,8){\circle{2}}
\put(22,16){\circle{2}}


\put(0,8){\line(1,0){01}} \put(0,12){\line(1,0){01}}
\put(0,16){\line(1,0){01}}

\put(2,12){\circle{2}} \put(2,16){\circle{2}}
\put(2,8){\circle{2}}


\put(3,26){\line(0,-1){04}} \put(23,26){\line(0,-1){04}}
\put(53,26){\line(0,-1){04}} \put(73,26){\line(0,-1){04}}

\put(3,26){\line(1,0){70}}

\put(04,22){\makebox(0,8)[bl]{\(X^{1}\) }}
\put(24,22){\makebox(0,8)[bl]{\(X^{2}\) }}

\put(34,20){\makebox(0,8)[bl]{{\bf .~ .~ .} }}

\put(54,22){\makebox(0,8)[bl]{\(X^{n-1}\) }}
\put(74,22){\makebox(0,8)[bl]{\(X^{n}\) }}


\put(07,26){\line(0,1){14}} \put(07,40){\circle*{3}}

\put(10,40){\makebox(0,8)[bl]{System: \(S = X^{1} \star X^{2}
 \star ... \star X^{n-1} \star X^{n} \) }}

\put(10,36){\makebox(0,8)[bl]{Example of system composition }}

\put(10,32){\makebox(0,8)[bl]{(configuration): }}

 \put(22,28){\makebox(0,8)[bl]{\(S_{1}=X^{1}_{2}\star
 X^{2}_{1} \star ... \star X^{n-1}_{2} \star X^{n}_{3}\)}}

\put(83,37){\makebox(0,8)[bl]{{\it Layer of}}}
\put(83,34){\makebox(0,8)[bl]{{\it system}}}

\put(83,25){\makebox(0,8)[bl]{{\it Layer of} }}
\put(83,22){\makebox(0,8)[bl]{{\it components/}}}
\put(83,19.5){\makebox(0,8)[bl]{{\it modules}}}

\put(83,13){\makebox(0,8)[bl]{{\it Layer of}}}
\put(83,10){\makebox(0,8)[bl]{{\it DAs}}}

\end{picture}
\end{center}

 The system composition problem can be based on
 multiple choice problem
 or morphological clique problem
 (while taking into account compatibility between the selected
 DAs) \cite{lev98,lev06,lev09,lev12morph,lev12a,lev15}.
 Fig. 5 illustrates the system composition
 for a four-component system
 while taking into account compatibility of DAs
 (concentric presentation).

\begin{center}
\begin{picture}(110,60)
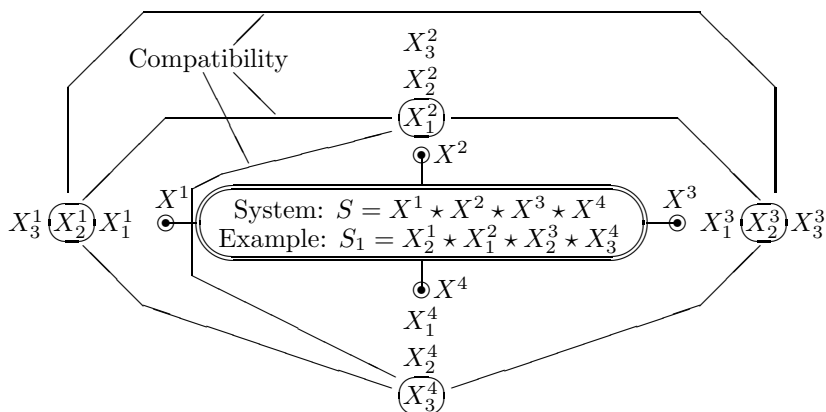
\put(010,00){\makebox(0,0)[bl] {Fig. 5. Concentric presentation of
 system composition}}

\put(16,50){\makebox(0,0)[bl]{Compatibility}}

\put(30,53.7){\line(1,1){04}} \put(30,50){\line(1,-1){05.5}}

\put(26,50){\line(1,-2){06}}


\put(08,34){\line(0,1){14}} \put(08,48){\line(1,1){10}}
\put(18,58){\line(1,0){74}}

\put(102,48){\line(0,-1){14}} \put(92,58){\line(1,-1){10}}

\put(00,28){\makebox(0,8)[bl]{\(X^{1}_{3}\)}}
\put(06,28){\makebox(0,8)[bl]{\(X^{1}_{2}\)}}
\put(12,28){\makebox(0,8)[bl]{\(X^{1}_{1}\)}}

\put(08.5,30){\oval(6,5)}

\put(21,30){\circle{2}} \put(21,30){\circle*{1}}
\put(19.5,32){\makebox(0,8)[bl]{\(X^{1}\)}}

\put(25,30){\line(-1,0){4}}

\put(24.5,23){\line(0,1){11.5}} \put(24.5,34.5){\line(2,1){4}}
\put(28.5,36.5){\line(4,1){22.5}} \put(24.5,23){\line(2,-1){27}}


\put(104,28){\makebox(0,8)[bl]{\(X^{3}_{3}\)}}
\put(98,28){\makebox(0,8)[bl]{\(X^{3}_{2}\)}}
\put(92,28){\makebox(0,8)[bl]{\(X^{3}_{1}\)}}

\put(100.5,30){\oval(6,5)}

\put(89,30){\circle{2}} \put(89,30){\circle*{1}}
\put(87,32){\makebox(0,8)[bl]{\(X^{3}\)}}

\put(85,30){\line(1,0){4}}


\put(59,44){\line(1,0){30}} \put(89,44){\line(1,-1){11}}

\put(51,44){\line(-1,0){30}} \put(21,44){\line(-1,-1){11}}

\put(52.5,52){\makebox(0,8)[bl]{\(X^{2}_{3}\)}}
\put(52.5,47){\makebox(0,8)[bl]{\(X^{2}_{2}\)}}
\put(52.5,42){\makebox(0,8)[bl]{\(X^{2}_{1}\)}}

\put(55,44){\oval(6,5)}

\put(55,39){\circle{2}} \put(55,39){\circle*{1}}
\put(56.5,38){\makebox(0,8)[bl]{\(X^{2}\)}}

\put(55,35){\line(0,1){4}}


\put(55,30){\oval(60,10)} \put(55,30){\oval(59,09)}

\put(30,30){\makebox(0,8)[bl]{System: \(S = X^{1} \star X^{2}
  \star X^{3} \star X^{4} \) }}

\put(28,26){\makebox(0,8)[bl]{Example: \(S_{1}=X^{1}_{2}\star
 X^{2}_{1}  \star X^{3}_{2} \star X^{4}_{3}\)}}

\put(55,25){\line(0,-1){4}}

\put(55,21){\circle{2}} \put(55,21){\circle*{1}}
\put(56.5,20){\makebox(0,8)[bl]{\(X^{4}\)}}

\put(52.5,15){\makebox(0,8)[bl]{\(X^{4}_{1}\)}}
\put(52.5,10){\makebox(0,8)[bl]{\(X^{4}_{2}\)}}
\put(52.5,05){\makebox(0,8)[bl]{\(X^{4}_{3}\)}}

\put(55,07){\oval(6,5)}

\put(59,08){\line(3,1){33}} \put(92,19){\line(1,1){08}}
\put(51,08){\line(-3,1){33}} \put(18,19){\line(-1,1){08}}

\end{picture}
\end{center}

  In the case of DAs,
 the following information
 is considered (i.e., morphological system structure)
 (e.g., \cite{lev98,lev06,lev09,lev12hier,lev15}:
 (a) estimates of DAs
  (e.g., vector estimates, ordinal estimates,
   interval multiset estimates),
 (b) estimates of compatibility between DAs of different system components
 (e.g., ordinal estimates, interval multiset estimates).

 Further, two illustrations are presented:
 (i) hierarchical (tree-like) system model (Fig. 6) and
 (ii) hierarchical system model with common modules
 for subsystems (Fig. 7).


\begin{center}
\begin{picture}(90,44)

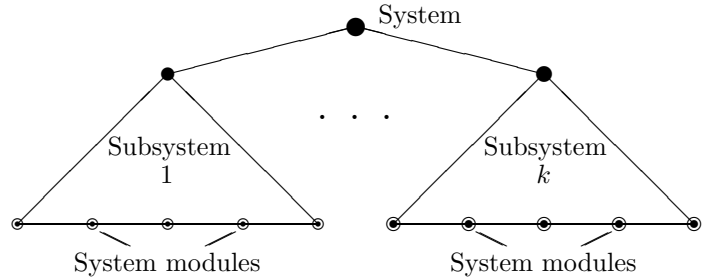
\put(10,00){\makebox(0,0)[bl]{Fig. 6. Hierarchical
 (tree-like) system model }}

\put(48,39){\makebox(0,0)[bl]{System}}

\put(45,39.25){\circle*{2.4}}

\put(20,33){\line(4,1){25}} \put(70,33){\line(-4,1){25}}


\put(12,21.5){\makebox(0,0)[bl]{Subsystem}}
\put(19,18.5){\makebox(0,0)[bl]{\(1\)}}

\put(20,33){\circle*{1.8}}

\put(00,13){\line(1,0){40}} \put(00,13){\line(1,1){20}}
\put(40,13){\line(-1,1){20}}

\put(00,13){\circle{1.5}} \put(00,13){\circle*{0.8}}
\put(10,13){\circle{1.5}} \put(10,13){\circle*{0.8}}
\put(20,13){\circle{1.5}} \put(20,13){\circle*{0.8}}
\put(30,13){\circle{1.5}} \put(30,13){\circle*{0.8}}
\put(40,13){\circle{1.5}} \put(40,13){\circle*{0.8}}

\put(11,12){\line(2,-1){4}} \put(29,12){\line(-2,-1){04}}

\put(07.5,06){\makebox(0,0)[bl]{System modules}}


\put(40,27){\makebox(0,0)[bl]{{\bf . ~. ~.}}}


\put(62,21.5){\makebox(0,0)[bl]{Subsystem}}
\put(69,18.5){\makebox(0,0)[bl]{\(k\)}}

\put(70,33){\circle*{2}}

\put(50,13){\line(1,0){40}} \put(50,13){\line(1,1){20}}
\put(90,13){\line(-1,1){20}}

\put(50,13){\circle{1.8}} \put(50,13){\circle*{1.2}}
\put(60,13){\circle{1.8}} \put(60,13){\circle*{1.2}}
\put(70,13){\circle{1.8}} \put(70,13){\circle*{1.2}}
\put(80,13){\circle{1.8}} \put(80,13){\circle*{1.2}}
\put(90,13){\circle{1.8}} \put(90,13){\circle*{1.2}}

\put(61,12){\line(2,-1){4}} \put(79,12){\line(-2,-1){04}}

\put(57.5,06){\makebox(0,0)[bl]{System modules}}

\end{picture}
\end{center}

\begin{center}
\begin{picture}(70,41)

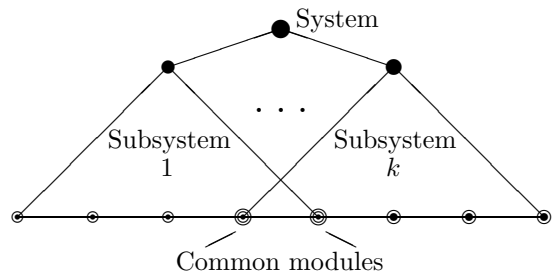
\put(03.5,00){\makebox(0,0)[bl]{Fig. 7. Common modules
 for subsystem}}

\put(37,37.6){\makebox(0,0)[bl]{System}}

\put(35,38){\circle*{2.4}}

\put(20,33){\line(3,1){15}} \put(50,33){\line(-3,1){15}}


\put(12,21.5){\makebox(0,0)[bl]{Subsystem}}
\put(19,18.5){\makebox(0,0)[bl]{\(1\)}}

\put(20,33){\circle*{1.8}}

\put(00,13){\line(1,0){40}} \put(00,13){\line(1,1){20}}
\put(40,13){\line(-1,1){20}}

\put(00,13){\circle{1.5}} \put(00,13){\circle*{0.8}}
\put(10,13){\circle{1.5}} \put(10,13){\circle*{0.8}}
\put(20,13){\circle{1.5}} \put(20,13){\circle*{0.8}}

\put(30,13){\circle{2}} \put(30,13){\circle{1.5}}
\put(30,13){\circle*{0.8}}

\put(40,13){\circle{2}} \put(40,13){\circle{1.5}}
\put(40,13){\circle*{0.8}}

\put(29,11){\line(-2,-1){4}} \put(41,11){\line(2,-1){04}}

\put(21,06){\makebox(0,0)[bl]{Common modules}}


\put(31.5,27){\makebox(0,0)[bl]{{\bf . . .}}}


\put(42,21.5){\makebox(0,0)[bl]{Subsystem}}
\put(49,18.5){\makebox(0,0)[bl]{\(k\)}}

\put(50,33){\circle*{2}}

\put(30,13){\line(1,0){40}} \put(30,13){\line(1,1){20}}
\put(70,13){\line(-1,1){20}}

\put(50,13){\circle{1.7}} \put(50,13){\circle*{1.0}}
\put(60,13){\circle{1.7}} \put(60,13){\circle*{1.0}}
\put(70,13){\circle{1.7}} \put(70,13){\circle*{1.0}}

\end{picture}
\end{center}

\section{Decision Problem Trends from Alternative To Composite Alternative}

 Main decision problem trends in moving process from
 alternative(s) to composite alternative(s)
 (i.e., composite systems) is depicted in Fig. 8
  \cite{lev98,lev06,levprob07,lev13intro,lev15}.

\begin{center}
\begin{picture}(115,78)

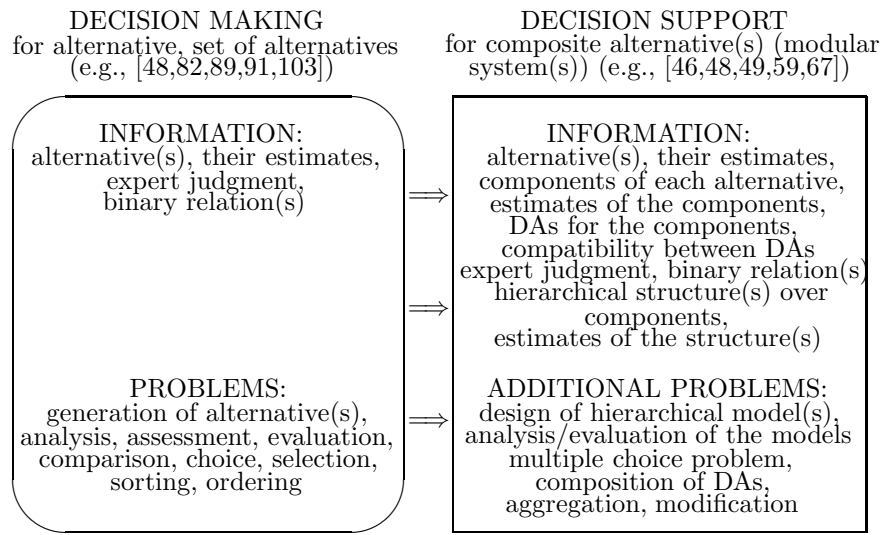
\put(01,00){\makebox(0,0)[bl]{Fig. 8. Moving from
 alternative(s) to composite (modular) alternative(s)}}


\put(09,73){\makebox(0,0)[bl]{DECISION MAKING}}
\put(01.5,69){\makebox(0,0)[bl]{for alternative, set of
 alternatives}}

\put(09,66){\makebox(0,0)[bl]{(e.g.,
 \cite{lev06,mirkin05,roy96,sim58,zop02})}}


\put(27.5,35){\oval(52,58)}


\put(013,58){\makebox(0,0)[bl]{INFORMATION:}}
\put(04,54){\makebox(0,0)[bl]{alternative(s), their estimates,}}
\put(014,51){\makebox(0,0)[bl]{expert judgment, }}

\put(013.5,48){\makebox(0,0)[bl]{binary relation(s)}}

\put(017,24){\makebox(0,0)[bl]{PROBLEMS:}}
\put(06,20){\makebox(0,0)[bl]{generation of alternative(s),}}
\put(02.5,17){\makebox(0,0)[bl]{analysis, assessment,
 evaluation,}}

\put(05,14){\makebox(0,0)[bl]{comparison, choice, selection,}}

\put(014.5,11){\makebox(0,0)[bl]{sorting, ordering}}

\put(54,50){\makebox(0,0)[bl]{\(\Longrightarrow\)}}
\put(54,35){\makebox(0,0)[bl]{\(\Longrightarrow\)}}
\put(54,20){\makebox(0,0)[bl]{\(\Longrightarrow\)}}


\put(69,73){\makebox(0,0)[bl]{DECISION SUPPORT}}

\put(59,69){\makebox(0,0)[bl]{for composite alternative(s)
(modular}}

\put(61,66){\makebox(0,0)[bl]{ system(s)) (e.g.,
 \cite{lev98,lev06,levprob07,lev13intro,lev15})}}

\put(60,06){\line(1,0){55}} \put(60,64){\line(1,0){55}}
\put(60,06){\line(0,1){58}}\put(115,06){\line(0,1){58}}


\put(72.5,58){\makebox(0,0)[bl]{INFORMATION:}}
\put(64.5,54){\makebox(0,0)[bl]{alternative(s), their estimates,}}
\put(63.5,51){\makebox(0,0)[bl]{components of each alternative,}}
\put(65,48){\makebox(0,0)[bl]{estimates of the components,}}
\put(67.5,45){\makebox(0,0)[bl]{DAs for the components,}}
\put(66.5,42){\makebox(0,0)[bl]{compatibility between DAs}}
\put(61,39){\makebox(0,0)[bl]{expert judgment, binary
 relation(s)}}

\put(65.5,36){\makebox(0,0)[bl]{hierarchical structure(s) over}}
\put(77.5,33){\makebox(0,0)[bl]{components,}}
\put(65.5,30){\makebox(0,0)[bl]{estimates of the structure(s)}}

\put(64.5,24){\makebox(0,0)[bl]{ADDITIONAL PROBLEMS:}}
\put(63.5,20){\makebox(0,0)[bl]{design of hierarchical model(s),}}
\put(61.6,17){\makebox(0,0)[bl]{analysis/evaluation of the
 models}}

\put(67.5,14){\makebox(0,0)[bl]{multiple choice problem,}}

\put(71,11){\makebox(0,0)[bl]{composition of DAs,}}

\put(67,08){\makebox(0,0)[bl]{aggregation, modification}}

\end{picture}
\end{center}

 Evidently, the decision problems became to be
 more complicated by several directions, for example:

 (a) hierarchical structures (models) of composite alternatives and their processing
  (design of hierarchical strcuture/model, evaluation,
  comparison, modification, aggregation);

 (b) components of each composite alternative and DAs for each
  component (including assessment and evaluation of the DAs),
  assessment and evaluation of compatibility between DAs for
  alternative components.

 In addition, it is reasonable to point out
 basic types of resources and corresponding
 kinds of resource requirements
 (i.e., human resources, information-computing requirements
 (Fig. 9).

\begin{center}
\begin{picture}(115,79)

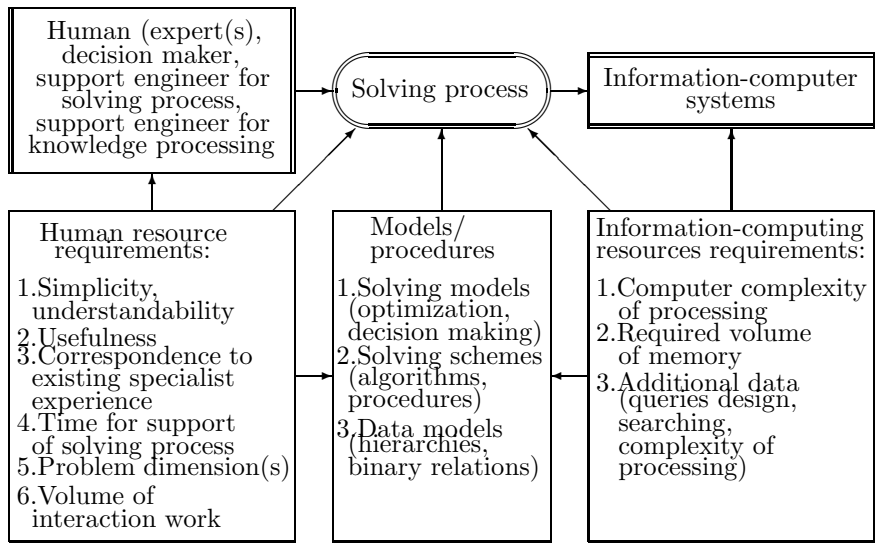
\put(26,00){\makebox(0,0)[bl]{Fig. 9. Scheme of resource
  requirements}}

\put(38,66){\vector(1,0){5}} \put(72,66){\vector(1,0){5}}


\put(45.5,64.5){\makebox(0,0)[bl]{Solving process}}


\put(57.5,66){\oval(29,10)} \put(57.5,66){\oval(28,09)}


\put(00,55){\line(1,0){38}} \put(00,77){\line(1,0){38}}
\put(00,55){\line(0,1){22}}\put(38,55){\line(0,1){22}}
\put(0.4,55){\line(0,1){22}}\put(37.5,55){\line(0,1){22}}

\put(05,72){\makebox(0,0)[bl]{Human (expert(s),}}
\put(07,69){\makebox(0,0)[bl]{decision maker, }}
\put(03.6,66){\makebox(0,0)[bl]{support engineer for}}
\put(07,63){\makebox(0,0)[bl]{solving process,}}
\put(03.6,60){\makebox(0,0)[bl]{support engineer for }}
\put(02.5,57){\makebox(0,0)[bl]{knowledge processing}}


\put(77,61){\line(1,0){38}} \put(77,71){\line(1,0){38}}
\put(77,61.5){\line(1,0){38}} \put(77,70.5){\line(1,0){38}}

\put(77,61){\line(0,1){10}}\put(115,61){\line(0,1){10}}

\put(79,66){\makebox(0,0)[bl]{Information-computer}}
\put(90,63){\makebox(0,0)[bl]{systems}}


\put(19,50){\vector(0,1){5}} \put(96,50){\vector(0,1){11}}
\put(35,50){\vector(1,1){11}} \put(80,50){\vector(-1,1){11}}
\put(57.5,50){\vector(0,1){11}}


\put(00,06){\line(1,0){38}} \put(00,50){\line(1,0){38}}
\put(00,06){\line(0,1){44}}\put(38,06){\line(0,1){44}}

\put(04,46){\makebox(0,0)[bl]{Human resource}}
\put(06,43){\makebox(0,0)[bl]{requirements:}}

\put(01,38){\makebox(0,0)[bl]{1.Simplicity,}}
\put(03,35){\makebox(0,0)[bl]{understandability}}

\put(01,32){\makebox(0,0)[bl]{2.Usefulness}}

\put(01,29){\makebox(0,0)[bl]{3.Correspondence to}}
\put(03,26){\makebox(0,0)[bl]{existing specialist}}
\put(03,23){\makebox(0,0)[bl]{experience}}

\put(01,20){\makebox(0,0)[bl]{4.Time for support}}
\put(03,17){\makebox(0,0)[bl]{of solving process}}

\put(01,14){\makebox(0,0)[bl]{5.Problem dimension(s)}}
\put(01,11){\makebox(0,0)[bl]{6.Volume of}}
\put(03,08){\makebox(0,0)[bl]{interaction work}}


\put(38,28){\vector(1,0){5}} \put(77,28){\vector(-1,0){5}}


\put(43,06){\line(1,0){29}} \put(43,50){\line(1,0){29}}
\put(43,06){\line(0,1){44}} \put(72,06){\line(0,1){44}}

\put(48,46){\makebox(0,0)[bl]{Models/}}
\put(48,43){\makebox(0,0)[bl]{procedures}}

\put(43.5,38){\makebox(0,0)[bl]{1.Solving models}}
\put(45,35){\makebox(0,0)[bl]{(optimization,}}
\put(45,32){\makebox(0,0)[bl]{decision making)}}

\put(43.5,29){\makebox(0,0)[bl]{2.Solving schemes}}
\put(45,26){\makebox(0,0)[bl]{(algorithms, }}
\put(45,23){\makebox(0,0)[bl]{procedures)}}

\put(43.5,20){\makebox(0,0)[bl]{3.Data models }}
\put(45,17){\makebox(0,0)[bl]{(hierarchies,}}
\put(45,14){\makebox(0,0)[bl]{binary relations)}}


\put(77,06){\line(1,0){38}} \put(77,50){\line(1,0){38}}
\put(77,06){\line(0,1){44}}\put(115,06){\line(0,1){44}}

\put(78,46){\makebox(0,0)[bl]{Information-computing}}
\put(78,43){\makebox(0,0)[bl]{resources requirements:}}

\put(78,38){\makebox(0,0)[bl]{1.Computer complexity}}
\put(81,35){\makebox(0,0)[bl]{of processing}}

\put(78,32){\makebox(0,0)[bl]{2.Required volume}}
\put(81,29){\makebox(0,0)[bl]{of memory}}

\put(78,26){\makebox(0,0)[bl]{3.Additional data}}
\put(81,23){\makebox(0,0)[bl]{(queries design,}}
\put(81,20){\makebox(0,0)[bl]{searching,}}
\put(81,17){\makebox(0,0)[bl]{complexity of}}
\put(81,14){\makebox(0,0)[bl]{processing)}}


\end{picture}
\end{center}

\section{Support Problems/Frameworks and Applications}

 Seven support combinatorial engineering framework for modular systems (composite alternatives)
 have been suggested by the author
 (e.g., \cite{lev06,levprob07,lev13intro}):

 1. Morphological system design (combinatorial synthesis)
 based on hierarchical multicriteria morphological design (HMMD) approach
  (an hierarchical extension of morphological analysis
  while taking into account
  ordinal estimates of DAs and their compatibility)
 \cite{lev98,lev06,lev12morph,lev12a,lev15}.

 2. Design of hierarchical system models
  (i.e., tree-like structures)
  \cite{lev12hier,lev15}:

 3. Evaluation of hierarchical modular system
   \cite{lev06,lev13eva,lev15}.

 4. Detection of system bottlenecks
 \cite{lev12clique,lev13bot,lev15}.

 5. System improvement/extension
 \cite{lev98,lev06,lev13imp,lev15}.


 6. Multistage system design (design of system trajectory)
 \cite{lev13tra,lev15}.

 7. Combinatorial system evolution and forecasting
 \cite{lev13,lev15,lev09had,levand10}.

 Table 2 contains some applied examples  for the combinatorial engineering frameworks
 above.


 In the case of grouping the application examples
  by large discipline domains (Fig. 3),
  the following groups for application examples are obtained:

 {\it 1.} Engineering domains:
 control engineering
 (management system for smart homes)
 \cite{lev13home,lev15},
 communication engineering
 (GSM system, standard for multimedia information processing)
 \cite{lev13mpeg,lev15,lev09had},
 protocol engineering (communication protocol ZigBee)
 \cite{lev12zig,lev13mpeg,lev15,levand10},
 sensor/telemetry systems \cite{lev15,levkhod07,levfim10},
 civil engineering (building from the viewpoint of earthquake engineering)
 \cite{lev06,levdan05}.
%

 {\it 2.} Computer science:
 software engineering \cite{lev06},
 information systems \cite{lev98},
  configuration of
 applied Web-based information systems \cite{lev11inf,lev15},
 composite retrieval \cite{lev98,lev12shop,lev15}.

  {\it 3.} Management, planning:
   geological planning \cite{lev98},
   investment \cite{lev98},
   medical treatment \cite{lev06,lev15,levsok04}.

  {\it 4.} Life cycle engineering/management:
 concrete technology (design, manufacturing, transportation,
 utilization \cite{lev06,levnis01}.

%
  {\it 5.} Education (engineering, applied mathematics, CS):
 design and combinatorial modeling of courses on system design
 \cite{lev98,lev06,lev15}.

~~

~~

~~

~~

~~

~~

~~

~~

\begin{center}
{\bf Table 2.} Examples of applications (combinatorial engineering frameworks)\\
\begin{tabular}{| l  l  | l |}
\hline

  & Support engineering framework & Some application(s) \\
\hline

 1. &Combinatorial synthesis&
  Modular software \cite{lev05,lev06} \\
  &&Management system in smart home \cite{lev13home,lev15}\\
 &&GSM communication network \cite{lev12morph,lev15} \\
  &&Wireless sensor element \cite{lev15,levfim10}\\
 &&On-board telemetry system \cite{lev15,levkhod07}\\
 &&Medical treatment \cite{lev06,lev15,levsok04}\\
   &&Vibration conveyor \cite{lev98}\\
     &&Concrete technology \cite{lev06,levnis01}\\
      &&Immunoassay technology \cite{lev06,levfir05}\\
 &&Web-based information system \cite{lev11inf,lev15}\\
 &&Communication protocol ZigBee \cite{lev12zig,lev13mpeg,lev15} \\
 &&Standard for multimedia   \cite{lev13mpeg,lev15,lev09had}\\
 &&Composite product in electronic shopping \cite{lev12shop,lev15}\\

 2. &Hierarchical system modeling&
    Management system in smart home \cite{lev13home,lev15}\\

  &&Communication protocol ZigBee \cite{lev12zig,lev13mpeg,lev15,levand10} \\
   &&Concrete technology \cite{lev06,levnis01}\\
   &&Immunoassay technology \cite{lev06,levfir05}\\
  &&Standard for multimedia   \cite{lev13mpeg,lev15,lev09had}\\
   &&On-board telemetry system \cite{lev15,levkhod07}\\
  &&Medical treatment \cite{lev06,lev15,levsok04}\\
   &&Vibration conveyor \cite{lev98}\\
  &&Wireless sensor element \cite{lev15,levfim10}\\
  &&Web-based information system \cite{lev11inf,lev15}\\
  &&Composite product in electronic shopping \cite{lev12shop,lev15}\\
   && Building \cite{lev06,levdan05}\\


 3. &Evaluation of  system&
 Composite product in electronic shopping \cite{lev12shop,lev15}\\

  &&Wireless sensor element \cite{lev15,levfim10}\\
    &&Vibration conveyor \cite{lev98}\\
     &&Concrete technology \cite{lev06,levnis01}\\
      &&Immunoassay technology \cite{lev06,levfir05}\\
   &&On-board telemetry system \cite{lev15,levkhod07}\\
  &&Management system in smart home \cite{lev13home,lev15}\\
  &&Communication protocol ZigBee \cite{lev12zig,lev13mpeg,lev15,levand10} \\
  &&Standard for multimedia   \cite{lev13mpeg,lev15}\\
  &&Medical treatment \cite{lev06,lev15,levsok04}\\
  &&Web-based information system \cite{lev11inf,lev15}\\
  && Building \cite{lev06,levdan05}\\

 4. &Detection of bottlenecks&
 Web-based information system \cite{lev11inf,lev15}\\

  &&On-board telemetry system \cite{lev15,levkhod07}\\
    &&Wireless sensor element \cite{lev15,levfim10}\\
 5. &System improvement/extension&
   Management system in smart home \cite{lev13home,lev15}\\

  &&On-board telemetry system \cite{lev15,levkhod07}\\
    &&Wireless sensor element \cite{lev15,levfim10}\\
 && Building \cite{lev06,levdan05}\\

 6.& Multistage  design&
   Modular education courses  \cite{lev13,lev15}\\

  && Web-based information system \cite{lev11inf,lev15}\\

 7. &Evolution and forecasting&
     Modular education courses  \cite{lev13,lev15}\\

    &&Standard for multimedia \cite{lev15,lev09had} \\
    &&Communication protocol ZigBee \cite{lev15,levand10} \\
    &&Web-based information system \cite{lev11inf,lev15}\\


\hline
\end{tabular}
\end{center}

\section{Conclusion}

 This  paper contains the author's glance to
 a general decision support platform technology for
 modular systems (i.e., composite/modular alternatives).
 Evidently, the decision support platform is an open system
 and can be extended, for example:
 (i) additional combinatorial optimization models (e.g., \cite{gar79,nem99,papa98,roberts76}),
 (ii) additional composite combinatorial frameworks (e.g., \cite{lev11a,lev15}).
  It is reasonable to point out the several future research
  directions for the described decision support platform:

 {\it 1.} the platform
 may be considered as a prospective tool for
 modular system design, evaluation, and maintenance;

 {\it 3.}  the  platform may be interesting from the viewpoint
 of new decision support systems for composite (modular)
 alternatives; and

 {\it 2.}  the platform
 is a significant direction for
 contemporary support systems
 in the field of system/product life cycle engineering/management.

\end{document}